\documentclass[usenatbib]{mn2e}
\usepackage{epsfig}
\def\gtwid{\mathrel{\raise.3ex\hbox{$>$\kern-.75em\lower1ex\hbox{$\sim$}}}}
\def\ltwid{\mathrel{\raise.3ex\hbox{$<$\kern-.75em\lower1ex\hbox{$\sim$}}}}
\def\\{\hfil\break}
\def\ie{{\it i.e.\ }}

\def\hmpc{$h^{-1}$Mpc}
\newcommand{\degree}{{\rm o}}
\newcommand{\nmom}{N$_{\rm MOM}$}

\newcommand{\mjh}[1]{}

\begin{document}

\title[Cosmic Flows Minimum Variance Moments]
{Cosmic Flows on 100$h^{-1}$ Mpc Scales:  Standardized Minimum Variance Bulk Flow, Shear and Octupole Moments}

\author[Feldman, Watkins \& Hudson]{Hume A. Feldman$^{\star,1,4,5}$, Richard Watkins$^{\dagger,2}$ \& Michael J. Hudson$^{\ddagger,3,6,7}$\\
$^1$Department of Physics and Astronomy, University of Kansas, Lawrence, KS 66045, USA.\\
$^2$Department of Physics, Willamette University, Salem, OR 97301, USA.\\
$^3$Department of Physics and Astronomy, University of Waterloo, Waterloo, ON N2L 3G1, Canada.\\
$^4$Department of Physics and Astronomy, University College London, London, WC1E 6BT, United Kingdom.\\
$^5$Department of Physics, Blackett Laboratory, Imperial College, London SW7 2AZ, United Kingdom.\\
$^6$Institut d'Astrophysique de Paris - UMR 7095, CNRS/Universit\'{e} Pierre et Marie Curie, 98bis boulevard Arago, 75014 Paris, France. \\
$^7$Perimeter Institute for Theoretical Physics, 31 Caroline St. N., Waterloo, ON, N2L 2Y5, Canada. \\
emails: $^\star$feldman@ku.edu;\,  $^\dagger$rwatkins@willamette.edu;\, $^\ddagger$mjhudson@uwaterloo.ca}

\maketitle

\begin{abstract}
The low order moments, such as the bulk flow and shear, of the large scale peculiar velocity field are sensitive probes of the matter density fluctuations on very large scales.   In practice, however, peculiar velocity surveys are usually sparse and noisy, which can lead to the aliasing of small scale power into what is meant to be a probe of the largest scales.  Previously, we developed an optimal ``minimum variance'' (MV) weighting scheme, designed to overcome this problem by minimizing the difference between the measured bulk flow (BF) and that which would be measured by an ideal survey.  Here we extend this MV analysis to include the shear and octupole moments, which are designed to have almost no correlations between them so that they are virtually orthogonal. We apply this MV analysis to a compilation of all major peculiar velocity surveys, consisting of 4536 measurements. Our estimate of the BF on scales of $\sim$ 100\hmpc\ has a magnitude of $|v|= 416 \pm 78$ km/s towards Galactic $l = 282^\degree \pm 11^\degree$ and $b =   6^\degree \pm  6^\degree$. This result is in disagreement with $\Lambda$CDM with WMAP5 cosmological parameters at a high confidence level, but is in good agreement with our previous MV result without an orthogonality constraint, showing that the shear and octupole moments did not contaminate the previous BF measurement. The shear and octupole moments are consistent with WMAP5 power spectrum, although the measurement noise is larger for these moments than for the BF.  The relatively low shear moments suggest that the sources responsible for the BF are at large distances.
\end{abstract}

\noindent{\it Subject headings}: cosmology: distance scales -- cosmology: large scale structure of the universe -- cosmology: observation -- cosmology: theory -- galaxies: kinematics and dynamics -- galaxies: statistics

\section{Introduction}
\label{sec:intro}





Large-scale structure formation is assumed to arise from small Gaussian initial fluctuations amplified by gravitational instability \citep{BBKS,EisHu98}. This basic framework is strongly supported by the consistency between the Cosmic Microwave Background (CMB) angular power spectra \citep[hereafter WMAP5]{WMAP5} observed at high-redshift, and large-scale structure (LSS) data from gravitational lensing \citep{FuSemHoe08} and galaxy power spectra \citep{EisZehHog05,Cole05} and its bispectrum \citep{bispec1,bispec2,Verde02}, measured  at low redshift.

However, on the largest scales, the comparisons are usually between the \emph{matter} density fluctuations measured by the CMB and the fluctuations in \emph{galaxy} number density, and so are susceptible to uncertainties in the relationship between mass and light (``biasing'').  While dark matter can be observed directly at relatively low redshift via gravitational lensing, at present this technique only just reaches into the linear regime \citep{FuSemHoe08}. The Integrated Sachs-Wolfe (ISW) effect \citep{SacWol67} probes the evolution of the dark matter potential on large scales via a cross-correlaton between galaxies and the CMB. Recent ISW compilations suggest a stronger signal than expected \citep{HoHirPad08}. Perhaps the most promising method for probing the low-redshift DM power spectrum on very large scales ($\ltwid100$\hmpc, where $h$ is the Hubble constant in units of 100 km s$^{-1}$ Mpc $^{-1}$)  is via the peculiar velocity field of galaxies and clusters \citep{StrWil95}. 

Peculiar velocities can be measured statistically through redshift-space distortions of galaxy power spectra \citep{PeaColNor01,TegBlaStr04}, or more directly by measuring distances to individual galaxies or clusters.  Recent peculiar velocity studies, using various techniques \citep{PikHud05, ParPar06, SarFelWat07, WatFel07, FelWat08, WatFelHud09}, all suggest that different peculiar velocity surveys, using different distance estimators, are all consistent with sampling the same underlying peculiar velocity field. Moreover, recent peculiar velocity surveys \citep{sfi1,sfi2,sfierr} are deeper, denser, and more reliable than ever before.  As surveys have gotten larger, our understanding of the distance indicators needed to extract the peculiar velocities, and to control their systematic errors, has also improved. Finally, new analytic techniques have allowed us to better extract information from surveys \citep{pairwisef, RadLucHud04,PikHud05, SarFelWat07, WatFel07, LavTulMoh10}.

On small scales ($\ltwid 20$ \hmpc), peculiar velocities yield results consistent with the WMAP5+BAO+SN $\Lambda$CDM model value: $(\Omega_{\mathrm m}/0.3)^{0.55} \sigma_{8} = 0.77\pm 0.035$ \citep{KomDunNol09}.  For example, on these scales, \citet{PikHud05} found $(\Omega_{\mathrm m}/0.3)^{0.55} \sigma_{8} = 0.80\pm0.05$ from a comparison of density (galaxy redshift) and peculiar velocity surveys.  A statistical analysis of pairwise velocities \citep{pairwisef} yields, after correction for non-linearities \citep{sigma810}, a slightly higher result for $\sigma_{8}$: $1.02\pm0.18$, but which is still consistent with \cite{KomDunNol09}. \cite{AbaErd09} found  $(\Omega_{\mathrm m}/0.25)^{0.55} \sigma_{8} = 0.90^{+0.22}_{-0.16}$  from a correlation function analysis of SFI++ (Spiral Field I band) over a range of scales $\gtwid 25$ \hmpc.

On the very largest scales, however, the agreement is less clear. Previous work has suggested that much of the peculiar velocity of the Local Group (LG) and nearby galaxies is generated by gravitational sources on very large scales, at a level which may be in excess of expectations from the standard $\Lambda$CDM model. This evidence comes directly from estimates of the bulk flow (BF) on large scales \citep{FelWat08, WatFelHud09}. It also comes indirectly from galaxy redshift surveys which allow one to estimate how much of the LG's motion is generated locally: the measurement of a low amplitude, misaligned \emph{gravity} dipole from to masses within  $\gtwid 60-100$ \hmpc\ then requires that a large component of the LG's motion is generated externally, on larger scales \citep{PikHud05, LavTulMoh10}.

Many previous studies of peculiar velocity data have been based on maximum-likelihood analysis of the entire velocity field\citep{ZarBerdaC01,AbaErd09}. While such studies have the advantage that they use all of the information present, there are disadvantages that arise because all scales are analyzed simultaneously. In particular they are sensitive to treatment of quasi- and non-linear regimes, and to the details of the assumed peculiar velocity errors.  Moreover, such studies typically assume a given power spectrum shape \emph{a priori}.  In this work, we isolate the largest scales.

In a recent paper (Watkins et al. 2009, hereafter Paper I), we developed the ``minimum variance'' (MV) moments that were designed to estimate the BF on a particular scale with minimal sensitivity to small-scale power. Paper I also showed that the BF's from independent peculiar velocity surveys were consistent with each other.   In this paper, we extend the formalism to include the next higher elements in the expansion, namely the shear and octupole moments. These higher-order moments contain information about the power spectrum on scales that are large, but not as large as that probed by the BF. The primary goal of this paper is to assess whether these moments, like the BF, have higher amplitude than expected than expected in $\Lambda$CDM. Moreover, they also allow us to extract cosmographical information: for example, \citet{LilYahJon86} first used the existence of a BF and a shear in the very nearby Universe to determine the approximate distance to the Great Attractor.  

It is important to note that in our analysis, since we are fitting to moments of an idealized survey, our model does not change with additional moments, that is, we are estimating individual moments rather than fit a model.   In principle, then, our estimates of bulk flow from Paper I should not change in the analysis presented here.   However, in practice, our use of an orthogonality constraint as described in Section~\ref{sec:mom} below will lead to small changes in our bulk flow estimates as we include higher moments.

In Section~\ref{sec:mom} we introduce the MV weights and the nineteen MV moments for the BF, shear and octupole. In Section~\ref{sec:data}, we discuss the peculiar velocity catalogues  analyzed here. In Section~\ref{sec:results} we present the moment amplitudes and compare these with expectations from cosmological models. We discuss our results in Section~\ref{sec:discussion} and conclude in Section~\ref{sec:conclusions}.

\section{Moments of the Velocity Field}
\label{sec:mom}

While large-scale flows are still in the linear regime, the statistics of individual galaxy or cluster peculiar velocities $S_n$ are not well described by linear theory due to the existence of nonlinear flows on small scales.    This problem may be solved by decomposing the velocity field into components, each of which reflects motions on a particular range of scales.   Since the statistics of moments associated with large-scale motions can be treated using linear theory, the amplitudes of these moments can be used to put direct constraints on cosmological parameters.

The most commonly used decomposition of the velocity field is a Taylor series expansion \citep{Kai88,JafKai95}, where the components of the velocity field are written as 
\begin{equation}
v_i ({\bf r}) = U_i + U_{ij}r_j + U_{ijk}r_jr_k+...
\end{equation}
The three zeroth order constants, $U_i$, are the components of the oft-discussed ``bulk flow" (BF).   The first order ``shear" tensor, $U_{ij}$, is symmetric if the velocity field is assumed to be curl-free, as it must be if motions are caused by gravitational instability.   Thus the condition that $U_{ij}=U_{ji}$ gives six independent shear components.   The second order tensor, $U_{ijk}$, which we will call the ``octupole" tensor, must also be symmetric for the same reason, so that there are ten independent octupole components, giving a total of 19 independent components for a second-order expansion.

This expansion is used to decompose the velocity field in a particular volume, typically that occupied by the galaxies in a peculiar velocity survey.   Under these circumstances, the BF moments probe scales larger than the diameter of the volume, with each subsequent order probing smaller and smaller scales.   However, peculiar velocity surveys typically have complicated geometries,  so that moment amplitudes are not comparable between surveys \citep[see][]{WatFelHud09,SarFelWat07}.  Indeed, even the interpretation of these moments can be difficult due not only to the complicated distribution of survey objects but also to the varying measurement errors associated with each object.   


Interpretation of peculiar velocity data would be much more straightforward if the data were close to an ``ideal survey'': an infinitely large spherically symmetric survey with no measurement errors and with a gaussian radial distribution function $f(r)\propto e^{-r^2/2R^2_I}$, where the parameter $R_I$ designates the depth of the survey.   The velocity moments obtained in this way correspond to a known scale, are straightforward to interpret, and are comparable between surveys.   While we do not have such ideal surveys, we have some flexibility to force our actual surveys to match such an ideal survey as closely as possible. This flexibility is in the form of a weight for each peculiar velocity datum,  which we are free to adjust in an optimal way.

Note that this approach of adjusting the weights to match a given ``ideal ''geometry is very different from the standard maximum likelihood (MLE) weights, which minimize the measurement error but do not account for the geometry of the survey.

Before we discuss the estimation of the velocity moments using velocity surveys, we must address a problem with the Taylor expansion that arises at second-order and beyond.   While the zeroth and first order moments are orthogonal, there is significant overlap between the BF and the octupole moments.   In particular, a pure octupole flow in a given volume $V$, $v_i = U_{ijk}r_ir_k$, contains a net BF in that volume given by $ \int_V U_{ijk}r_ir_k\  d^3r$.   This leads to a strong correlation between the BF and octupole moments.  

We can solve this problem by modifying the definition of the octupole moments.   We rewrite our expansion of the velocity field in a volume $V$ as 
\begin{equation}
v_i ({\bf r}) = U_i + U_{ij}r_j + U_{ijk}\left(r_jr_k-\Lambda_{jk}\right)+...
\end{equation}
where the constants $\Lambda_{jk}$ are given by 
\begin{equation}
\Lambda_{jk} = \int_V r_jr_k\ d^3r
\end{equation}
For a spherically symmetric volume, as we use here, only the diagonal elements $\Lambda_{11}$, $\Lambda_{22}$, and $\Lambda_{33}$ are nonzero.   When the octupole moments are defined in this way, the 19 moments to second order are orthogonal and have no overlap.  A similar procedure can be carried out to remove overlap between the third order moments and the shear, however, current surveys are not sensitive to higher moments than the octupole and thus in this paper we will focus on the second order expansion only.   

Now that we have defined the velocity moments we are interested in, we turn to how these moments can be estimated using peculiar velocity data.  First, it is only possible to measure the line-of-sight peculiar velocity $s= \vec v\cdot \hat r$.   Our expansion for the peculiar velocity field thus translates into an expansion for the line-of-sight velocity field $s({\bf r})$, which can be written as
\begin{equation}
s({\bf r}) = U_i\hat r_i + U_{ij}r\hat r_i\hat r_j + U_{ijk}\left(r^2\hat r_i\hat r_j\hat r_k - \Lambda_{jk}\hat r_i\right) + ...
\label{eq:mtaylor}
\end{equation}
For simplicity, we follow \citet{JafKai95} and write the second-order expansion in the form of a 19 component vector of moment amplitudes,
\begin{equation}
s({\bf r}) = \sum_{p=1}^{19}U_pg_p({\bf r})
\end{equation}
where $U_p$ are the 19 moment amplitudes given by
\begin{eqnarray}
U_p&=&\{U_x,U_y,U_z,U_{xx},U_{yy},U_{zz},U_{xy},U_{yz},U_{zx},\\ \nonumber
&&U_{xxx},U_{yyy},U_{zzz}, U_{xxy},U_{yyz},U_{zzx},U_{xyy},\\ \nonumber
&&U_{yzz},U_{zxx},U_{xyz}\}
\label{eq:mom-amp}
\end{eqnarray}
and the mode functions are given by
\begin{eqnarray}
 g_p({\bf r}) &=&\{ \hat r_x, \hat r_y, \hat r_z, r\hat r_x^2, r\hat r_y^2, r\hat r_z^2, 2r\hat r_x\hat r_y, 2r\hat r_y\hat r_z,\\ \nonumber
 && 2r\hat r_x\hat r_z,r^2\hat r_x^3-\Lambda_{xx}\hat r_x, r^2\hat r_y^3-\Lambda_{yy}\hat r_y,\\ \nonumber
&&r^2\hat r_z^3-\Lambda_{zz}\hat r_z,3r^2\hat r_x^2\hat r_y-\Lambda_{xx}\hat r_y,
 \\ \nonumber
&&3r^2\hat r_y^2\hat r_z-\Lambda_{yy}\hat r_z, 3r^2\hat r_z^2\hat r_x-\Lambda_{zz}\hat r_x,\\ \nonumber
&&3r^2\hat r_y^2\hat r_x-\Lambda_{yy}\hat r_x,3r^2\hat r_z^2\hat r_y-\Lambda_{zz}\hat r_y,\\ \nonumber
&&3r^2\hat r_x^2\hat r_z-\Lambda_{xx}\hat r_z,6r^2\hat r_x\hat r_y\hat r_z\}
\label{eq:mode-fn}
\end{eqnarray}
where we have used the fact that only the diagonal elements of $\Lambda_{ij}$ are nonzero.  

We first consider an idealized survey, consisting of positions ${\bf r}_n$ and exact line-of-sight velocities $s_n$ for a spherically symmetric distribution of $N_o$ objects with a given distribution function $f(r)$.   In this case, the ideal velocity moments $U_p$ are just the projections of the velocities onto the mode functions, 
\begin{equation}
U_p = \frac1{N_o}\sum_{n=1}^{N_o} g_p({\bf r}_n)s_n
\label{eq:iup}
\end{equation}
Thus the moment amplitudes take the form of linear combinations of the velocities $\sum_n w^\prime_{p,n}s_n$ with the numerical values of the MV weights given by 
\begin{equation}
w^\prime_{p,n}= g_p({\bf r}_n)/N_o
\label{eq:iweight}
\end{equation}
We note here that  Eqn. (\ref{eq:iup}) applies only for moments that have no overlap, as is true for the moments defined in Eqn. (\ref{eq:mtaylor}).  The expression for moment amplitudes when overlapping moments are used is somewhat more complicated.

An actual peculiar velocity survey consists of $N$ objects with positions ${\bf r}_n$ and measured line-of-sight velocities $S_n$ with uncertainties $\sigma_n$.  The measured velocities are assumed to have the form $S_n= s_n + \delta_n$, where $\delta_n$ is drawn from a Gaussian distribution of variance $\sigma_n^2 + \sigma_*^2$.   Here $\sigma_*$ is the velocity noise, which accounts for small-scale motions not included in the measured moment.
In the peculiar velocity literature, the approach often taken is one where one fits a flow model (whether parametric flow model, e.g.\ an expansion over many  components, or using a ``template'' based on e.g.\ the gravity of a galaxy density field). In such cases, it is important that the flow model is complete on all scales and that the noise estimates and $\chi^{2}$ values are reasonable. That is not the approach taken here: instead we are estimating large-scale moments, which are expected to be close to orthogonal. Hence the value of $\sigma_*$ affects our moments only very weakly: it modifies the measurement noise and hence the weights. While it has the strongest effect on nearby galaxies for which the velocity errors are smallest, these same galaxies are strongly downweighted by our $R_{I} = 50$ \hmpc\ MV weighting scheme. Consequently, changing the value of $\sigma_*$ does not alter any of our moments significantly.

Given an idealized survey with velocity moments $U_p$, we wish to determine the weights $w_{p,n}$ such that the linear combinations
\begin{equation}
u_p= \sum_{n=1}^N w_{p,n}S_n
\label{eq:up}
\end{equation}
give the best possible estimates of the $U_p$.   Following our previous work (Paper I), we calculate the weights by minimizing the average variance $\langle (U_p-u_p)^2\rangle$.    When considering only the BF moments in Paper I, we included additional constraints that ensured that the estimators would give the correct amplitude for a pure BF velocity field.   Here we implement a more general set of constraints that is applicable to higher-order velocity moments.   

Suppose that the flow field consisted only of BF, shear, and octupole moments, so that the line-of-sight velocities at positions ${\bf r}_n$ took the form $s_n = \sum_p U_p g_p({\bf r}_n)$.   In order for the estimators to give the correct amplitudes for the velocity moments on average for this flow field, \ie $\langle u_p\rangle= U_p$, we require that
\begin{equation}
\sum_n w_{p,n} g_q({\bf r}_n) = \delta_{pq}
\label{eq:con}
\end{equation}
This set of constraints can be implemented by using Lagrange multipliers.   Thus we seek to minimize the quantity
\begin{equation}
\langle (U_p-u_p)^2\rangle +\sum_q \lambda_{pq}\left(\sum_n w_{p,n} g_q({\bf r}_n) - \delta_{pq}\right)
\end{equation}
Expanding out the first term, plugging in the expression for $u_p$ from Eqn. (\ref{eq:up}), we can write this expression in terms of the weights $w_{p,n}$,
\begin{eqnarray}
\langle U_p^2\rangle -\sum_n 2w_{p,n}\langle S_nU_p\rangle + \sum_{n,m} w_{p,n}w_{p,m}\langle S_nS_m\rangle\\ \nonumber
 + \sum_q \lambda_{pq}\left(\sum_n w_{p,n} g_q({\bf r}_n) - \delta_{pq}\right)
\end{eqnarray}

To find the weights that minimize this expression, we take the derivative with respect to $w_{p,n}$, set it equal to zero:
\begin{equation}
-2\langle S_nU_p\rangle + 2\sum_m w_{p,m}\langle S_nS_m\rangle + \sum_q \lambda_{pq} g_q({\bf r}_n)=0 
\end{equation}
and solve for the weights,
\begin{equation}
w_{p,n} = \sum_m G^{-1}_{nm}\left(\langle S_mU_p\rangle - {1\over 2}\sum_q \lambda_{pq} g_q({\bf r}_m)\right)
\label{eq:w}
\end{equation}
where $G$ is the covariance matrix of the individual measured velocities,  $G_{nm}\equiv \langle S_nS_m\rangle$.
The values of the Lagrange multipliers can be found by plugging Eqn. (\ref{eq:w}) into Eqn. (\ref{eq:con})  and solving for $\lambda_{pq}$,
\begin{equation}
\lambda_{pq} = \sum_l\left[M_{pl}^{-1}\left( \sum_{m,n} G_{nm}^{-1}\langle S_mU_l\rangle g_q({\bf r}_n)-\delta_{lq} \right)\right]
\end{equation}
where the matrix $M$ is given by
\begin{equation}
M_{pq} = {1\over 2}\sum_{n,m} G^{-1}_{nm}g_p({\bf r}_n)g_q({\bf r}_m)
\end{equation}

Eqn. (\ref{eq:w}) gives us a formula for calculating the MV weights in terms of the covariance matrix $G_{nm}= \langle S_nS_m\rangle$ and the correlation $\langle S_mU_p\rangle$, both of which can be calculated  given a power spectrum model.  Using the fact that $S_n= s_n+\delta_n$ as described above and that $s_n$ and $\delta_n$ are independent,  we can write the covariance matrix as 
\begin{eqnarray}
G_{mn} &=&  \langle s_m s_n\rangle
+ \delta_{mn}(\sigma_*^2 + \sigma_n^2)\\ \nonumber
&=&\langle {\bf \hat r}_m\cdot {\bf v}({\bf r}_m)\ \   {\bf\hat r}_n\cdot {\bf v}({\bf r}_n)\rangle
+ \delta_{mn}(\sigma_*^2 + \sigma_n^2).
 \label{eq:gmn}
\end{eqnarray}
In linear theory the first term can be expressed as an integral over the density power spectrum $P(k)$, 
\begin{equation}
\langle {\bf \hat r}_m\cdot {\bf v}({\bf r}_m)\ \   {\bf\hat r}_n\cdot {\bf v}({\bf r}_n)\rangle
=  {\Omega_{m}^{1.1}\over 2\pi^2}\int   dk\  P(k)f_{mn}(k)
\label{eq:vcov}
\end{equation}
where the function $f_{mn}(k)$ is the angle averaged window function 
 \begin{eqnarray}
 f_{mn}(k) &=& \int {d^2{\hat k}\over 4\pi} \left( {\bf \hat r}_m\cdot {\bf \hat k} \right)\left( {\bf \hat r}_n\cdot {\bf \hat k} \right) \\ \nonumber
 &&\times\exp \left(ik\ {\bf \hat k}\cdot ({\bf r}_m - {\bf r}_n)\right)
 \label{eq:fmn}
\end{eqnarray}

The correlation $\langle S_mU_p\rangle$ is calculated in a similar way.  We generate an ideal survey by selecting $N_o$ random positions ${\bf r}^\prime_{n^\prime}$ with the desired radial distribution function.    We can then write
\begin{equation}
\langle S_mU_p\rangle = \sum_{n^\prime} w^\prime_{pn^\prime}\langle s_ms_{n^\prime}\rangle
\end{equation} 
where the weights $ w^\prime_{pn^\prime}$ are the ideal weights given in Eqn. (\ref{eq:iweight}) and we have assumed that measurement errors are uncorrelated with velocities.  The correlation $\langle s_ms_{n^\prime}\rangle$ can be calculated in the same manner as the elements of the covariance matrix $G$ (Eq. \ref{eq:gmn}).

Once we have calculated the weights for the MV moments, it is straightforward to calculate their correlation matrix, 
\begin{eqnarray}
R_{pq}&=& \langle u_pu_q\rangle = \sum_{mn}w_{pm}w_{qn}\langle S_mS_n\rangle\\ \nonumber
&=&\sum_{mn}w_{pm}w_{qn}G_{mn}
 \label{eq:rpq}
\end{eqnarray}
The moment covariance matrix can be separated into two parts 
\begin{equation}
R_{pq}=R^{(v)}_{pq}+R^{(\epsilon)}_{pq}\ , 
\label{eq:Rpq}
\end{equation}
corresponding to the two terms in $G_{nm}$.  The second term represents the noise in the measurement of the moment
\begin{equation}
R^{(\epsilon)}_{pq} = \sum_n w_{pn}w_{qn}\left(\sigma_n^2 +\sigma_*^2\right)
\end{equation} 
The first term is due to actual motions of the objects in the survey and can be written as an integral over the density fluctuation power spectrum, 
\begin{equation}
R^{(v)}_{pq} = {\Omega_{m}^{1.1}\over 2\pi^2}\int   dk\  P(k){\cal W}^2_{pq}(k)
\label{eq:Rpqv}
\end{equation}
where the angle averaged tensor window function is 
\begin{equation}
{\cal W}^2_{pq} =  \sum_{m,n}w_{pm}w_{qn}f_{mn}(k)
\label{eq:wf}
\end{equation}
The diagonal elements ${\cal W}^2_{pp}$ are the window functions for the individual moments $u_p$.   The window function for a particular moment indicates which scales that moment probes.   It can also be compared to the window function for the ideal moment to see the particular scales on which the moments differ. 

In summary, our MV method is a two-step process: for each moment, determine the optimal MV weights via Eq. \ref{eq:w}, then use these weights to measure the moments themselves (Eq. \ref{eq:up}). The weights depend only weakly on the power spectra and nuisance parameters ($\sigma_{*}$), so the measured moments are robust. For a given moment we can also calculate its measurement variance (the diagonal elements of $R_{pq}^{(e)}$) and the cosmic variance (via $R_{pq}^{(v)}$).  Note that the cosmic variance is for the actual MV weights and not the ideal ones. The two variances then allow us to compare the actual moment amplitudes with expectations from cosmological models.

\section{Data}
\label{sec:data}

As in Paper I, the peculiar velocity data used here compiles all of the major peculiar velocity surveys published to date,  with the exception of the survey of \cite{LauPos94} which was found in Paper I to be inconsistent with other data sets. As in Paper I, we have removed outliers by using the predictions of the IRAS-PSCz (Point Source Catalogue Survey z=redshift) density field \citep[for more details see Paper I;][]{HudSmiLuc04,NeiHudCon07}.   Each individual survey has a characteristic MLE depth, defined as  $\sum r_n w_n / \sum w_n$ where the MLE weights are $w_n = 1/(\sigma_n^2 + \sigma_*^2)$.

This compilation, which we label ``COMPOSITE'', consists of a number of surveys, the largest of which is the SFI++ peculiar velocity survey of spirals in the field and in groups \citep{sfi1,sfi2,sfierr}.  Here we use the data from the corrected dataset \citep{sfierr} rather than from the erroneous one \citep{sfi2}.  After rejecting about 1\% of the data, the sample consists of 2720 TF galaxies and 736 groups to make 3456 data points with characteristic depth of 35 \hmpc.  The SFI++ sample is all-sky, except for the Galactic plane ($|b| \ltwid 15$). As in Paper I, we find that all of the surveys we studied are consistent with each other, with the possible exception of the \cite{LauPos94} (LP) BCG survey.  The MV-weighted bulk flow of LP disagrees with that of the COMPOSITE catalogue on all scales. The level of disagreement on the larger scales,  corresponds to 99\% CL.   There are independent reasons to question the LP results: \cite{HudSmiLuc04} compared, cluster by cluster, the distance to the brightest cluster galaxy derived by LP to that derived from the FP for other cluster members, and found that in a few cases, these distances differed significantly, in the sense that the LP BCG distance was too large (i.e. the BCG was fainter than expected). They found that all of the discrepant BCGs for which HST images were available showed strong evidence for dust. For these reasons, we have chosen to reject LP from the COMPOSITE catalogues.

It is interesting to compare the results from SFI++ with a completely independent peculiar velocity dataset of comparable depth and statistical power. The consistency between these two catalogues, which use different data and methods, is an important check on our results. As in Paper I, we define  the ``DEEP'' compilation including 103 SNIa \citep{TonSchBar03}, 70 SC Tully-Fisher (TF) clusters \citep{GioHaySal98, DalGioHay99}, 56 SMAC fundamental plane (FP) clusters \citep{HudSmiLuc99, HudSmiLuc04}, 50 EFAR FP clusters \citep{ColSagBur01} and 15 TF clusters  \citep{Wil99b}. The DEEP catalogue consists of 294 data points, but because these are clusters or SNe their peculiar velocity errors are lower per object than for SFI++. The DEEP  compilation covers the whole sky outside the Galactic plane and has a characteristic depth of 50 \hmpc. 

In Paper I, we analyzed a further compilation, dubbed ``SHALLOW'' that consisted of the ENEAR \citep{daCBerAlo00, BerAlodaC02b, WegBerWil03} survey and a surface brightness fluctuations (SBF) survey \citep{TonDreBla01}. Because for our purposes this compilation is rather shallow we do not analyze it separately here, but include it in COMPOSITE.

In Paper I, we showed that all of these subsets (SFI++, DEEP and SHALLOW) are consistent with the same underlying velocity field, given their sparse spatial sampling. The formalism for comparing surveys with different window functions is given in Paper I (see Eq. (24) and Table 3). 

In summary, the COMPOSITE catalogue covers the whole sky outside the Galactic plane and has a characteristic depth of 34 \hmpc. It is based on 4536 peculiar velocity measurements, making it the largest peculiar velocity catalogue compiled and studied to date.

\section{Results}
\label{sec:results}

\begin{figure*}
     \includegraphics[width= 0.5\textwidth]{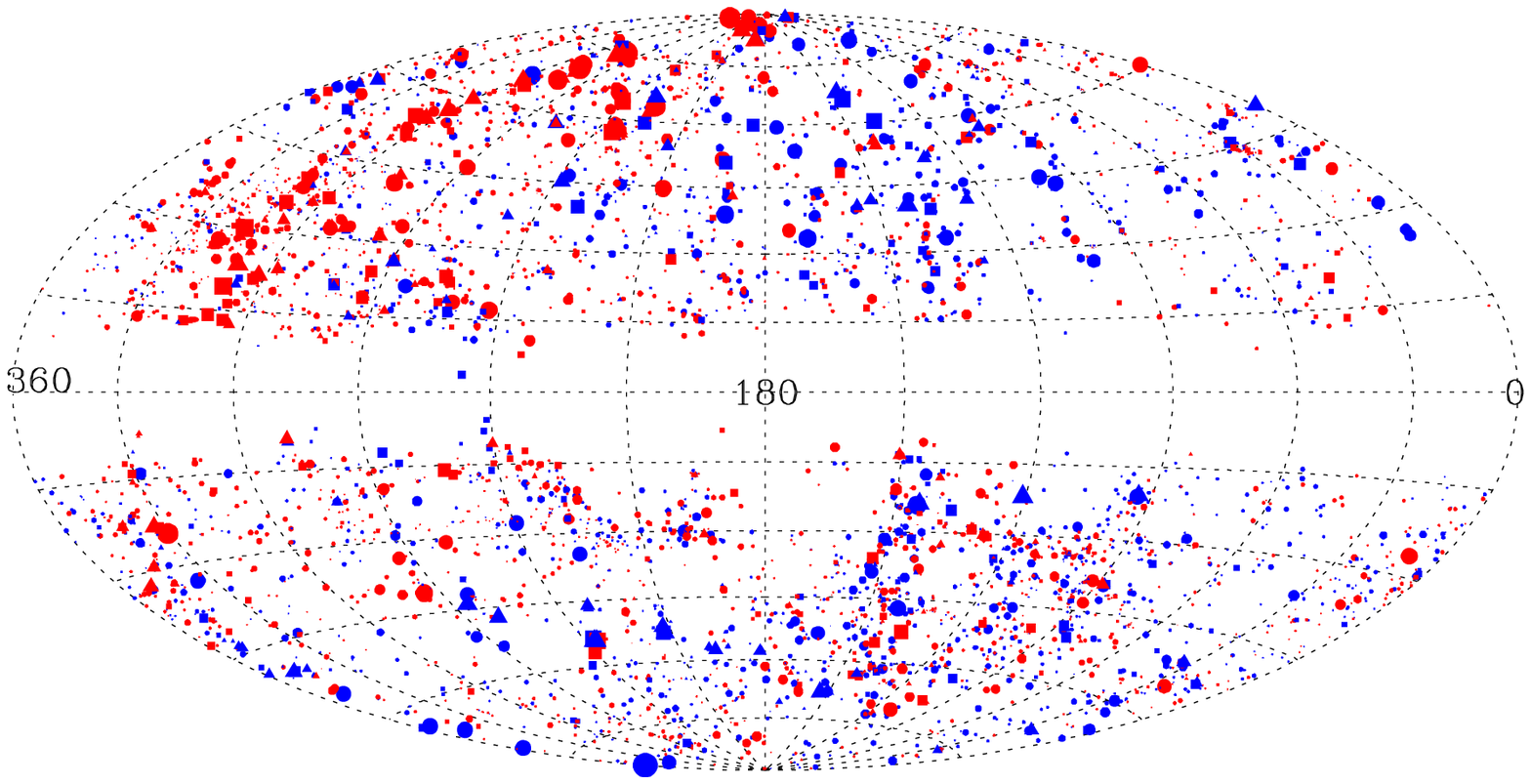}\includegraphics[width= 0.5\textwidth]{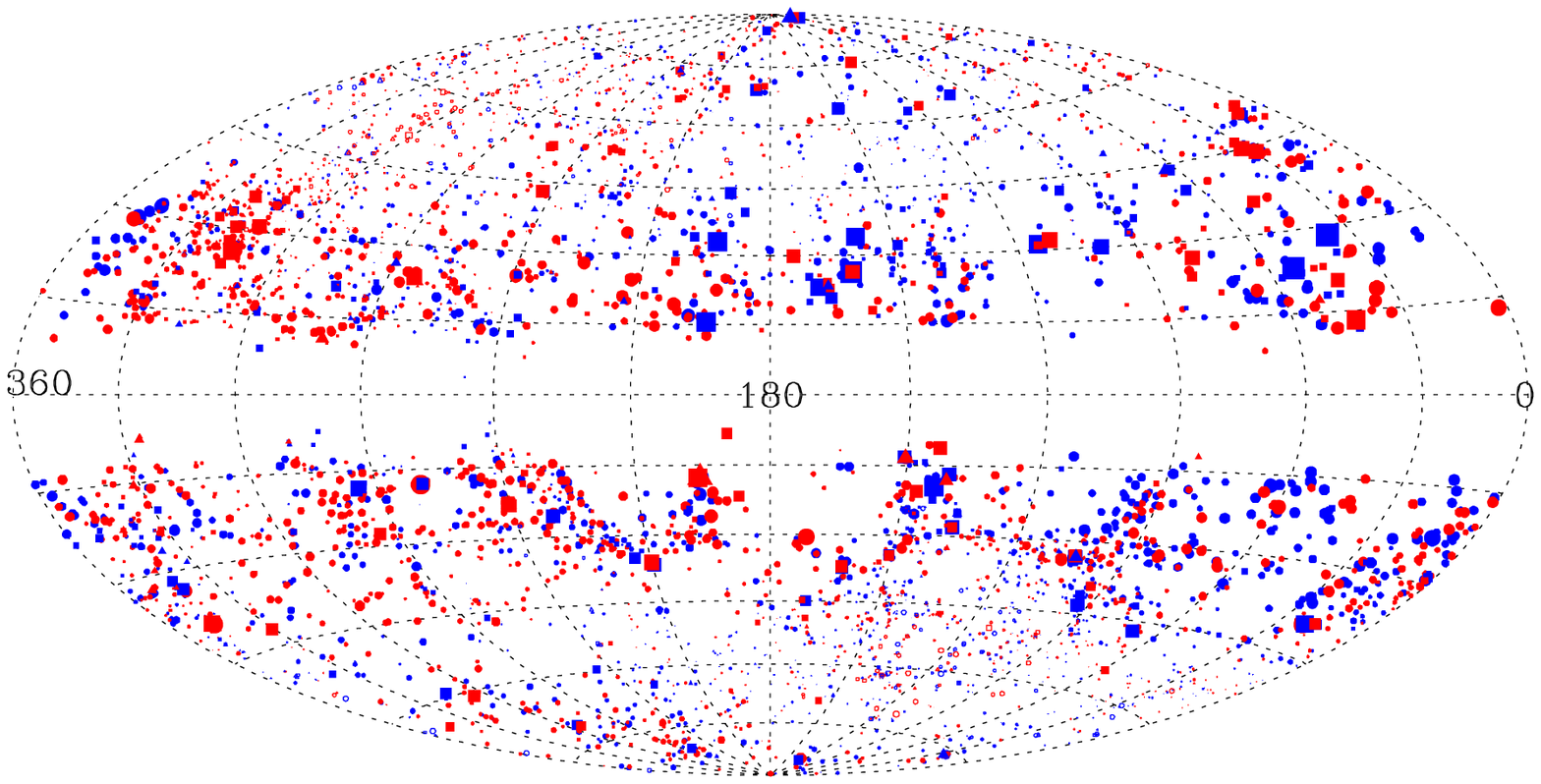}
        \caption{The COMPOSITE peculiar velocity catalogue on the sky (Galactic coordinates).  Points are colour coded by peculiar velocity with red outgoing and blue infalling. While it is customary in such plots to encode the amplitude of the peculiar velocity by the size of the symbol, in this case symbol area is proportional to the weight.  Data from SFI++ are shown by circles, DEEP by squares and SHALLOW by triangles. The left panel shows the MLE weights, whereas the right panel shows the average MV weights.}
    \label{fig:aitoff}
\end{figure*}

\subsection{Bulk Flow, Shear and Octupole Moments}
We calculated the MV moments for the BF, shear and octupole components using the methods described in Sec.~\ref{sec:mom} above for each of the catalogues described in Sec.~\ref{sec:data}.  For specificity, we used the $\Lambda$CDM power spectrum model of \cite{EisHu98} with the WMAP5 central parameters, $\sigma_8=0.796$, $h=0.719$, $\Omega_m=0.258$ and $n_s=0.963$ for the amplitude of cosmological density fluctuations, the Hubble constant, the normalized matter density and the spectral index, respectively and the velocity noise $\sigma_*=150$ km/s. The exact values of the cosmological parameters, including $\sigma_*$, make little difference to the values of the weights. 

In principle, our MV weights allow us to match any choice of ``ideal window'', which here is assumed to have a Gaussian profile, parametrized by $R_{I}$. Clearly, the larger, denser and more geometrically complete the catalogue is and the smaller the velocity errors are, the closer its window functions resemble that of the ideal window. In practice, however, there is a compromise between two competing goals: one is the need to adjust the weights, as best we can, so that the weighted catalogue matches the ideal survey, the other is to keep the noise small by down-weighting objects with the large measurement errors. Although we have investigated a range of scales $R_{I} \in[10,60]$ \hmpc, in this paper, we will focus on the window with $R_I=50$ \hmpc. This choice is a compromise between the desire to probe the largest possible scales, and the natural characteristic depths of the catalogues ($\sim 35$ \hmpc). The observational noise is minimized for low $R_{I}$ and becomes too large if $R_I > 50$ \hmpc. 

\begin{figure}
     \includegraphics[width=\columnwidth]{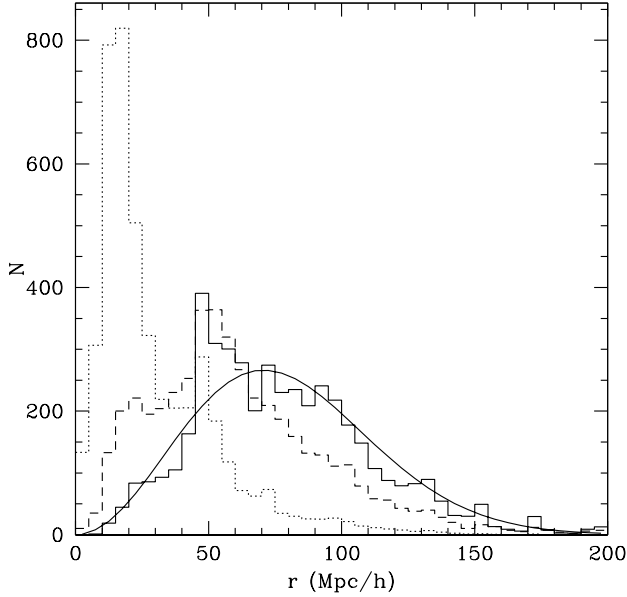}
        \caption{The COMPOSITE peculiar velocity catalogue as a function of depth. The dashed histogram shows the distribution of peculiar velocity measurements. The dotted histogram shows the \emph{weighted} histogram (renormalized to the same area) but using the usual MLE weights. Notice that most of the signal is driven by very nearby objects. The solid histogram is also weighted, but using MV weights. The smooth curve shows the expected weighted radial distribution for an ideal survey i.e. $\propto r^2  \exp\left[-r^2/(2 R_I^2)\right]$. This shows that our MV weighting scheme produces the desired radial distribution.}
    \label{fig:radial}
\end{figure}

Figure \ref{fig:aitoff} shows the peculiar velocity data on the sky, both for MLE weights and for MV weights. Notice how the MV weights become larger in regions where the spatial sampling is poorer, such as close to the Galactic plane. The weighted radial distribution galaxies is shown in Figure \ref{fig:radial}, where it is compared to the unweighted, MLE and $R_{I} = 50$ \hmpc\ ideal distributions. It is clear that the MV distribution closely approximates that of the ideal one.

\begin{figure}
     \includegraphics[width= \columnwidth]{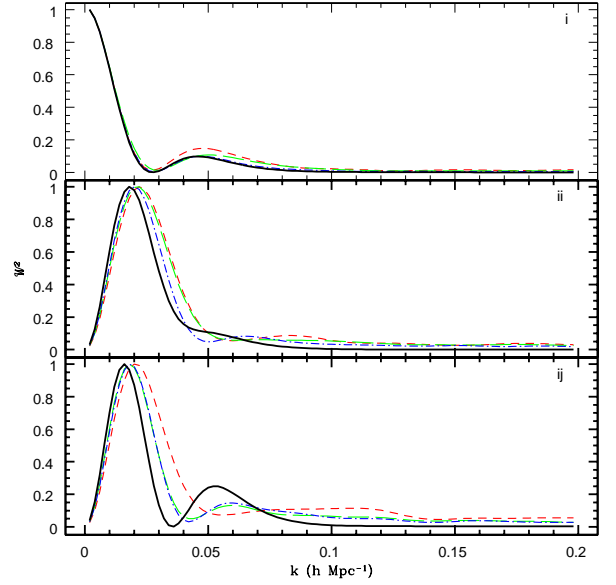}
        \caption{The window functions of the BF and shear moments for $R_I=50$\hmpc\  for the COMPOSITE catalogue. The thick  black lines are the ideal window functions for the MV  components (since the ideal survey is isotropic, all component are the same) whereas the thin lines are the actual components for the survey. In the top panel are the BF $x$ (short-dash), $y$ (long-dash), $z$ (dash-dot)--components. The middle panel shows the shear $xx$ (short-dash), $yy$ (long-dash), $zz$ (dash-dot) components, the bottom panel shows the shear $xy$ (short-dash), $yz$ (long-dash), and $zx$ (dash-dot) components. Notice how the WF virtually vanish at large $k$ (small scales). All coordinates are Galactic.}
    \label{fig:wf-bf-sh}
\end{figure}

\begin{figure}
     \includegraphics[width= \columnwidth]{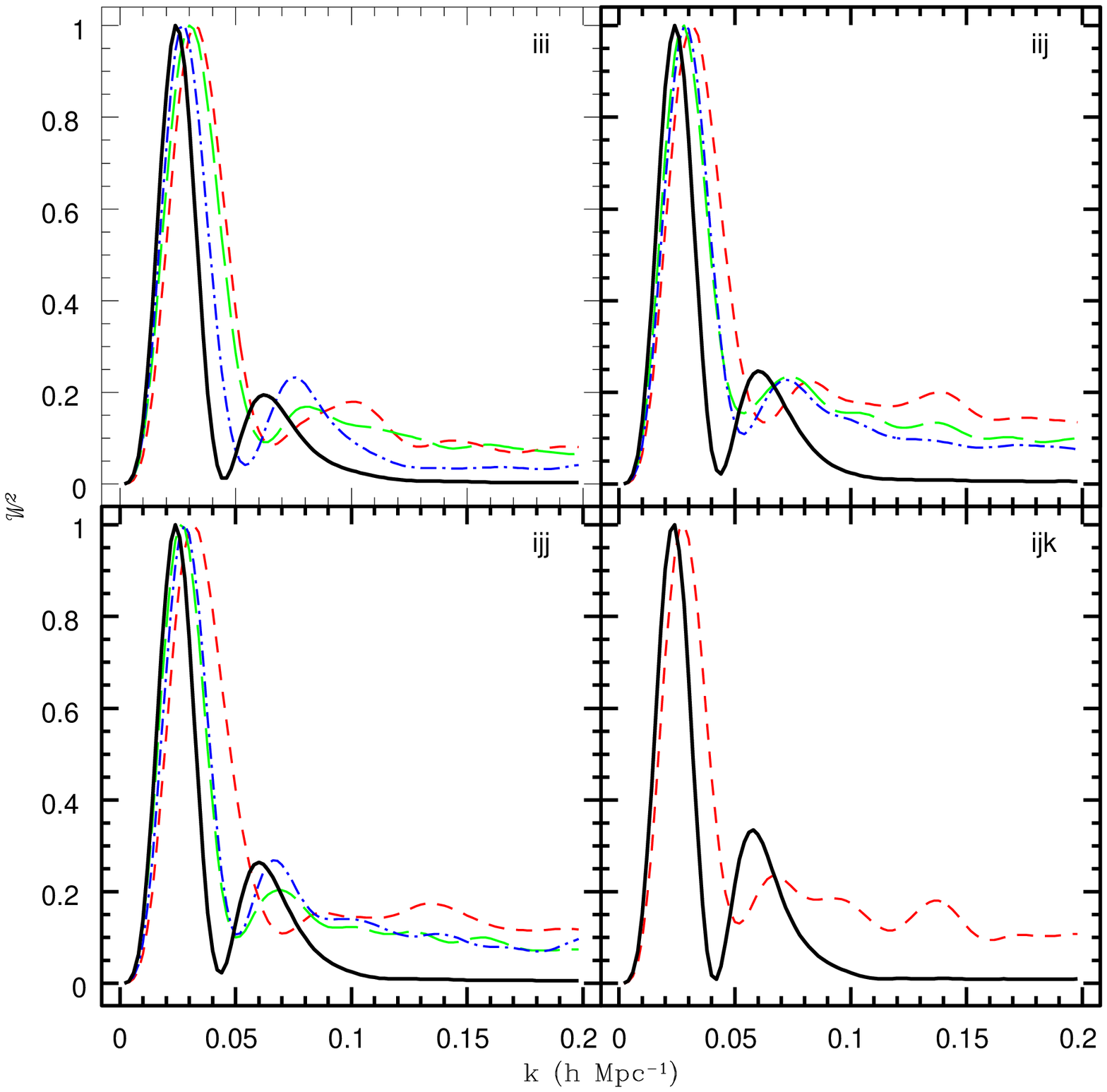}
        \caption{The same as \ref{fig:wf-bf-sh} for the octupole moments.  The thick  black lines represent the ideal WF. In the top left panel are the $xxx$ (short-dash), $yyy$ (long-dash), $zzz$ (dash-dot); the top right panel shows the $xxy$ (short-dash), $yyz$ (long-dash), $zzx$ (dash-dot), the bottom left panel shows the $xyy$ (short-dash), $yzz$ (long-dash), and $zxx$ (dash-dot); the bottom right panel shows the $xyz$ (short-dash) component of the WF's.}
    \label{fig:wf-octupole}
\end{figure}

We compared the window functions (WF) (Eq.~\ref{eq:wf}) of all surveys and composite catalogues to the ideal survey.  In Figures \ref{fig:wf-bf-sh} and \ref{fig:wf-octupole} we show both the ideal and survey MV BF, shear and octupole squared tensor window functions (Eq.~\ref{eq:wf}) for the COMPOSITE survey for $R_I=50$\hmpc. The advantage of the MV moments is that they have been designed to be sensitive {\it only} to a narrow range of scales, and so we will be able to probe these scales without having to worry about the influence of, or aliasing from, smaller scales. The BF moments probes scales much larger than $R_I$; the shear responds to $R_I$ scales and larger while the octupole to somewhat smaller scales (see Figures~\ref{fig:wf-bf-sh} and \ref{fig:wf-octupole}).  

When the WF's in Fig.~\ref{fig:wf-bf-sh} are convolved by the WMAP5 power spectrum, one finds that, for the BF statistic with $R_I=50$\hmpc, the contribution to the integrand in Eq.~\ref{eq:Rpqv} peaks at a wavenumber $k \sim 0.01$ h Mpc$^{-1}$, corresponding to wavelengths in excess of 600 \hmpc.  The shear is most sensitive to scales where the BF window is at its minimum : $k \sim 0.025$ h Mpc$^{-1}$,  or wavelengths of 250 \hmpc.  It is worth noting that these scales are similar to or larger than the very largest scales probed by Sloan Digital Sky Survey (SDSS) and the 2 degree Field Galaxy Redshift Survey (2dFGRS) \citep[e.g.][]{PerColEis07}. The octupole is sensitive to slightly smaller scales.

Figures  \ref{fig:wf-bf-sh} and \ref{fig:wf-octupole} also show that the ideal and survey window functions match well, especially for the lower moments. This is also apparent in Tables~\ref{tab:values-19}-\ref{tab:values-3} where we give the values of each moment and the correlation coefficient $\langle\frac{{\bf u}\cdot {\bf U}}{|{\bf u}||{\bf U}|}\rangle$ between the ideal and their MV estimates for each of the moments; a correlation coefficient of unity indicates perfect correlations. We see that the denser the catalogue, the better its match with the ideal moments.   

To compare the measured moments with expectations from cosmology, we need to need to know what moments would be expected. We have only a single measurement of each moment (because the flow field is expanded around the origin, i.e.\ at the location of the LG) but we can calculate what values might have been measured at other locations in a $\Lambda$CDM universe (assuming the Copernican principle).  For each moment we can determine its expected mean and variance by calculating ensemble averages over all possible observers.  Due to isotropy and homogeneity, the mean must vanish and so it is the variance that is the quantity of interest because it indicates the range of amplitudes for each moment that one would expect. We will call the corresponding standard deviation, the ``cosmic root mean square'' (CRMS) since it provides an estimate of the expected amplitude of the moment for a survey with the same geometry and weights, but with no measurement noise. 
In detail, the CRMS is given by the diagonal elements of the covariance matrix $R_{pq}^{(v)}$ (eq.~\ref{eq:Rpqv}), which in turn depends on the weights (and hence the measurement noise) and on the power spectrum, but does not include a contribution from the measurement noise.  Tables~\ref{tab:values-19}-\ref{tab:values-3} include for comparison the CRMS of each moment around its zero mean, given the spectrum specified by the WMAP5  central parameters and given the weights as determined above.   

\begin{table*}
\caption {The moment's value and its correlation coefficient $\langle\frac{{u}{U}}{|{u}||{U}|}\rangle$ for each of the catalogues and $R_I=50$\hmpc\ ; $N_{\rm MOM}=19$. All directions are given in Galactic coordinates. The last column is the expectation value of the moment given WMAP5 central parameters and the COMPOSITE catalogue WF's as described in the text.}
\begin{tabular}{ccc |||||| cc |||||| cc |||||| cc}
 & \multicolumn{2}{c}{  COMPOSITE} & \multicolumn{2}{c}{      SFI++} & \multicolumn{2}{c}{        DEEP}  & WMAP CRMS\\  
 \hline
   x &     86.5 $\pm$     68.8 &     0.74 &     69.0 $\pm$     95.7 &     0.64 &    192.7 $\pm$    115.6 &     0.51 & 110.6\\ \hline
   y &   -404.9 $\pm$     61.8 &     0.77 &   -473.6 $\pm$     87.2 &     0.67 &   -320.7 $\pm$    106.0 &     0.51  &   109.0\\ \hline 
   z &     42.8 $\pm$     37.7 &     0.89 &     57.7 $\pm$     59.3 &     0.80 &     62.0 $\pm$     55.8 &     0.76  & 105.3\\ \hline
  xx &      2.73 $\pm$      1.01 &     0.69 &      3.36 $\pm$      1.29 &     0.62 &      2.19 $\pm$      1.76 &     0.47 & 1.656 \\ \hline 
  yy &      1.37 $\pm$      0.98 &     0.69 &      3.72 $\pm$      1.27 &     0.63 &     -0.19 $\pm$      1.79 &     0.42  & 1.547 \\ \hline 
  zz &     -0.03 $\pm$      0.68 &     0.80 &      2.72 $\pm$      0.96 &     0.71 &     -0.72 $\pm$      1.04 &     0.67  &  1.462\\ \hline 
  xy &      0.13 $\pm$      0.76 &     0.51 &     -0.71 $\pm$      0.98 &     0.42 &      0.27 $\pm$      1.29 &     0.31 & 0.890\\ \hline  
  yz &     -0.95 $\pm$      0.57 &     0.63 &     -1.05 $\pm$      0.78 &     0.52 &     -0.71 $\pm$      0.94 &     0.40    &  0.749\\ \hline 
  zx &      1.22 $\pm$      0.54 &     0.66 &      1.50 $\pm$      0.74 &     0.56 &      0.98 $\pm$      0.84 &     0.47 &   0.767\\ \hline 
 xxx &    -1.2e-02 $\pm$     2.2e-02 &     0.38 &    -9.3e-03 $\pm$     2.9e-02 &     0.31 &     1.0e-02 $\pm$     3.6e-02 &     0.25 & 1.83e-02\\ \hline  
 yyy &    -2.4e-02 $\pm$     1.7e-02 &     0.41 &    -1.9e-02 $\pm$     2.4e-02 &     0.34 &    -2.2e-02 $\pm$     2.7e-02 &     0.24 & 1.43e-02 \\ \hline 
 zzz &    -7.2e-03 $\pm$     1.1e-02 &     0.61 &    -3.3e-03 $\pm$     1.6e-02 &     0.48 &    -2.5e-03 $\pm$     1.6e-02 &     0.47 & 1.28e-02 \\ \hline 
 xyy &    -8.2e-03 $\pm$     1.2e-02 &     0.30 &    -3.3e-02 $\pm$     1.7e-02 &     0.23 &     2.0e-02 $\pm$     1.9e-02 &     0.20 & 8.39e-03\\ \hline 
 yzz &     5.8e-04 $\pm$     6.6e-03 &     0.44 &    -1.8e-03 $\pm$     1.0e-02 &     0.33 &     8.9e-03 $\pm$     9.6e-03 &     0.30 & 5.44e-03 \\ \hline 
 zxx &     7.3e-03 $\pm$     7.8e-03 &     0.45 &     8.7e-03 $\pm$     1.1e-02 &     0.34 &    -2.1e-03 $\pm$     1.2e-02 &     0.34  & 6.60e-03  \\ \hline 
 xxy &     8.3e-03 $\pm$     1.2e-02 &     0.29 &     5.7e-03 $\pm$     1.6e-02 &     0.24 &     2.2e-02 $\pm$     1.9e-02 &     0.16  &   8.24e-03\\ \hline 
 yyz &     6.3e-04 $\pm$     8.3e-03 &     0.40 &     7.7e-03 $\pm$     1.2e-02 &     0.28 &    -2.5e-03 $\pm$     1.2e-02 &     0.30  & 6.35e-03 \\ \hline 
 zzx &     1.2e-02 $\pm$     7.6e-03 &     0.46 &    -2.5e-03 $\pm$     1.1e-02 &     0.35 &     1.6e-02 $\pm$     1.1e-02 &     0.34 &  6.86e-03\\ \hline 
 xyz &     6.6e-03 $\pm$     5.5e-03 &     0.34 &     9.3e-03 $\pm$     8.2e-03 &     0.25 &     4.9e-03 $\pm$     8.2e-03 &     0.22  &  3.72e-03\\ 
\hline  \hline
\end{tabular} 
\label{tab:values-19}
\end{table*}
 
\begin{table*}
\caption {The same as Table~\ref{tab:values-19} for $N_{\rm MOM}=9$} 
\begin{tabular}{ccc |||||| cc |||||| cc |||||| cc}
& \multicolumn{2}{c}{  COMPOSITE} & \multicolumn{2}{c}{      SFI++} & \multicolumn{2}{c}{        DEEP} & WMAP CRMS\\  
 \hline
    x &    101.8 $\pm$     38.4 &     0.87 &     65.0 $\pm$     46.9 &     0.82 &    127.6 $\pm$     62.9 &     0.70  & 110.6 \\ \hline  
   y &   -362.2 $\pm$     39.4 &     0.86 &   -361.6 $\pm$     47.8 &     0.81 &   -326.3 $\pm$     66.1 &     0.65   &  109.0\\ \hline 
   z &     40.1 $\pm$     30.9 &     0.92 &     76.6 $\pm$     40.3 &     0.87 &     49.2 $\pm$     47.8 &     0.80   &  105.3\\ \hline 
  xx &      2.89 $\pm$      0.98 &     0.69 &      3.94 $\pm$      1.25 &     0.63 &      2.39 $\pm$      1.61 &     0.51  & 1.656\\ \hline 
  yy &      1.21 $\pm$      0.95 &     0.69 &      3.90 $\pm$      1.22 &     0.64 &     -0.77 $\pm$      1.68 &     0.41  &  1.547\\ \hline 
  zz &      0.10 $\pm$      0.67 &     0.80 &      2.76 $\pm$      0.93 &     0.72 &     -0.69 $\pm$      1.00 &     0.68  & 1.462\\ \hline
  xy &      0.20 $\pm$      0.74 &     0.51 &     -0.76 $\pm$      0.95 &     0.43 &      0.22 $\pm$      1.23 &     0.31 & 0.890 \\ \hline 
  yz &     -1.02 $\pm$      0.55 &     0.63 &     -1.11 $\pm$      0.76 &     0.52 &     -0.63 $\pm$      0.86 &     0.41 &  0.749 \\ \hline  
  zx &      1.44 $\pm$      0.53 &     0.67 &      1.48 $\pm$      0.72 &     0.56 &      1.32 $\pm$      0.80 &     0.48 &   0.767\\ \hline
\hline
\end{tabular} 
\label{tab:values-9}
\end{table*}

\begin{table*}
\caption {The same as Table~\ref{tab:values-19} for $N_{\rm MOM}=3$} 
\begin{tabular}{ccc |||||| cc |||||| cc |||||| cc}
 & \multicolumn{2}{c}{  COMPOSITE} & \multicolumn{2}{c}{      SFI++} & \multicolumn{2}{c}{        DEEP} & WMAP CRMS\\  
 \hline
   x &     97.4 $\pm$     38.3 &     0.87 &     71.7 $\pm$     46.4 &     0.82 &    120.0 $\pm$     61.2 &     0.71   & 110.6\\ \hline
   y &   -361.9 $\pm$     39.2 &     0.86 &   -363.3 $\pm$     46.9 &     0.82 &   -333.8 $\pm$     65.4 &     0.65 &  109.0 \\ \hline
   z &     41.2 $\pm$     30.7 &     0.92 &     73.1 $\pm$     39.8 &     0.87 &     44.3 $\pm$     47.1 &     0.80   & 105.3 \\  \hline
\hline
\end{tabular} 
\label{tab:values-3}
\end{table*}

\begin{figure}
     \includegraphics[width= \columnwidth]{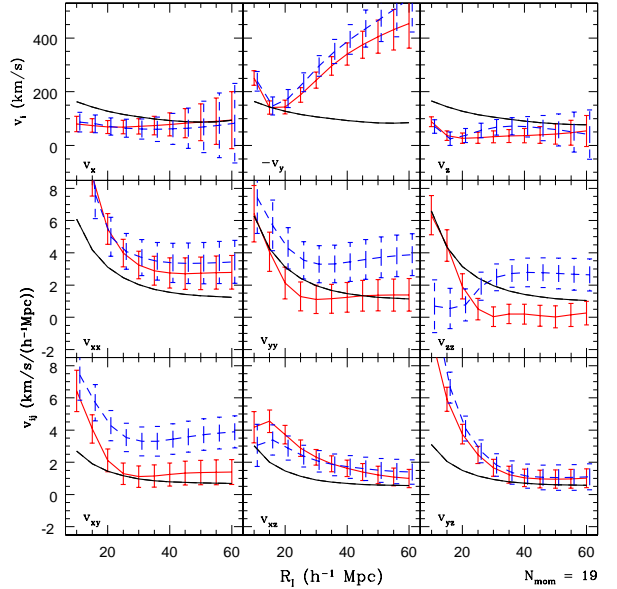}
        \caption{The BF and shear moments of the COMPOSITE (red solid line) and SFI++ (blue dashed line) catalogues as a function of $R_I$, all in Galactic coordinates. Note that the data points are not independent. The top panels are the BF velocities to the Galactic x (left), y (center) and z (left) directions. The error bars are as described in the text.   In the middle and bottom panels we show the shear components in the Galactic directions xx (middle left), yy (middle center) and zz (middle right) and the xy (bottom left) yz (bottom center and zx (bottom right). The solid black line, without the error bars, in each panel is the WMAP5 $\Lambda$CDM cosmic rms (CRMS).}
    \label{fig:flows-bf-sh}
\end{figure}

\begin{figure}
     \includegraphics[width= \columnwidth]{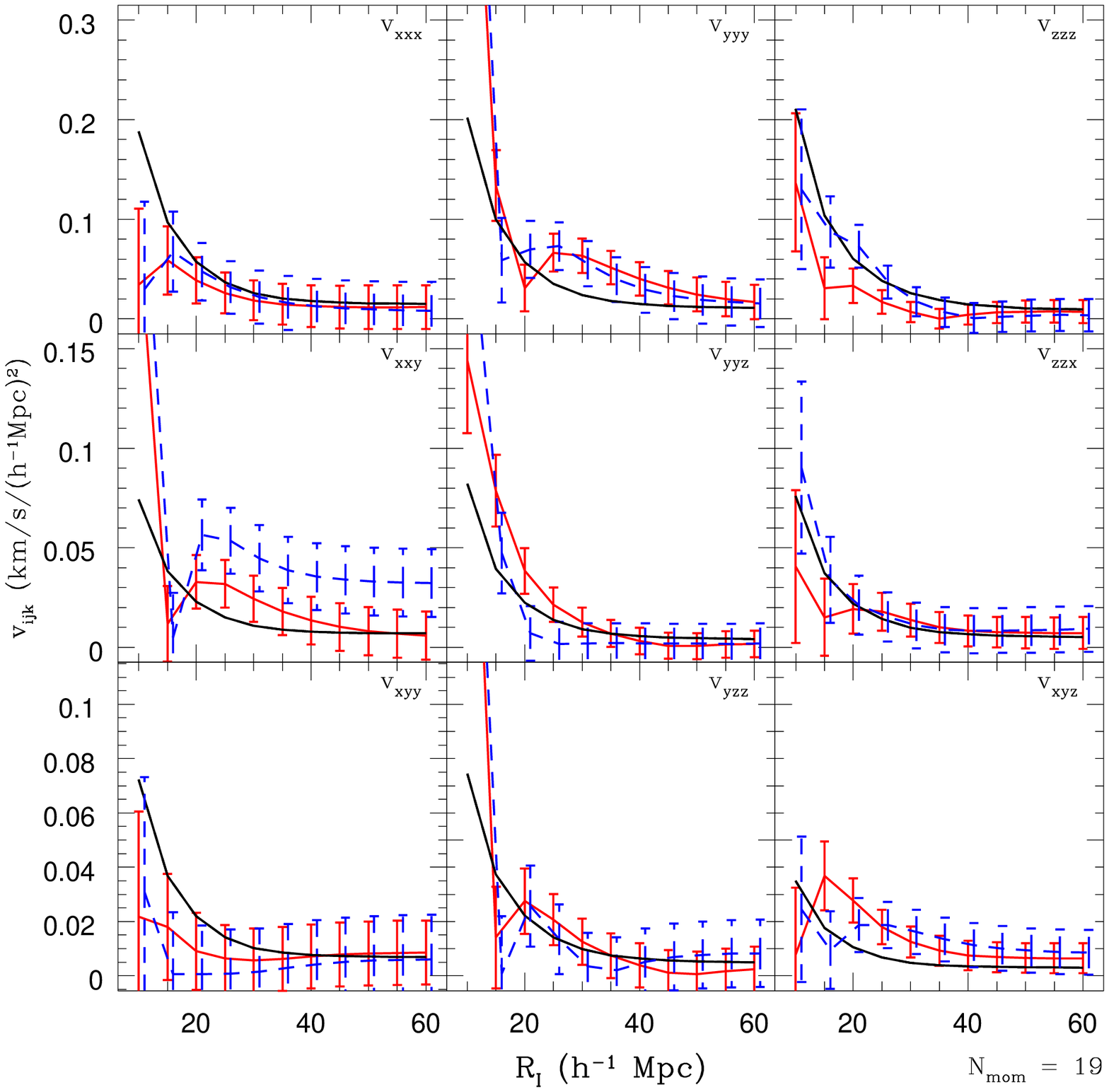}
        \caption{The same as Fig.~\ref{fig:flows-bf-sh} for the octupole moments. Galactic xxx (top left), yyy (top center) and zzz (top left), xxy (middle left), yyz (middle center) and zzx (middle right), the xyy (bottom left) yzz (bottom center and xyz (bottom right).}
    \label{fig:flows-octupole}
\end{figure}

Turning to the values of the moments themselves, Tables~\ref{tab:values-19}-\ref{tab:values-3} show that there is a remarkable agreement, within the errors, between all catalogues and also a good agreement with the CRMS (last column of the tables) for all moments except the Galactic y-direction component of the BF.  In Figures~\ref{fig:flows-bf-sh} and \ref{fig:flows-octupole} we show the bulk, shear and octupole flows as a function of the depth $R_I$ for the COMPOSITE and SFI++  surveys.  The error bars in Fig.  \ref{fig:flows-octupole} are for illustration and are the expected rms deviation, $\sqrt{\langle (U_i-u_i)^2\rangle}$, of the estimated moment from the actual value of the ideal moment. Thus they represent a combination of measurement noise and deviation from the ideal window. We also show the expected cosmic r.m.s.\ (CRMS) of the flow as described above. As noted above, the BF is particularly sensitive to the large scales of the matter power spectrum, and so the large amplitude of the BF is suggestive of excess power on scales $k \sim 0.01$ h Mpc$^{-1}$. 

\subsection{Comparison with $\Lambda$CDM cosmology}

To make the comparisons with cosmological models more precise, we compare, in a frequentist sense, the observed BF moments with the cosmological expectation.  In other words, we calculate the probability that a randomly placed observer would have observed moments as large as the ones measured. Specifically, since each moment is Gaussian-distributed around zero mean, we calculate the $\chi^2$ for \nmom\ degrees of freedom corresponding to the number of moments, as given by
\begin{equation}
\chi^2 = \sum_{p=1}^{N_{\rm MOM}}\sum_{q=1}^{N_{\rm MOM}} u_p R_{pq}^{-1}u_q,
\label{eq:chi}
\end{equation}
where $p$ and $q$ specify the BF, shear and octupole components of the covariance matrix $R_{pq}$ for a specific set of values for the cosmological parameters we are interested in. 

The results are presented in Table~\ref{tab:Pchisq}, where we show, for each of the catalogues, the percent probability of getting a larger $\chi^2$, $P(>\chi^2)$. In the table, we break down the probabilities for \nmom=3 (BF alone), \nmom=9 (total, BF and shear) and for \nmom=19 into total, BF, shear and octupole. This has the benefit of showing us clearly which part of the flow is in agreement or disagreement with standard cosmology. Table~\ref{tab:Pchisq} shows that although the BF is highly unlikely, both the shear and octupole moments amplitudes are more or less what is expected.  For \nmom=19, the total probability of getting the observed $\chi^2$ or higher is $\sim20$\%, the probability of the shear alone is $\in[10,50]$\% and the octupole $\sim80$\%, whereas getting the BF is $\ltwid 0.5$\%.

Note that the BF is discrepant at the 99.78\% and 97\% levels for the two independent subsamples (SFI++ and DEEP respectively).  Recall that the SFI++ peculiar velocities are based on the Tully-Fisher relation applied to field and group spirals, whereas the DEEP compilation is based on a variety of other methods, particularly Type Ia supernovae and the Fundamental Plane relation in early-type galaxies.  The agreement between these independent datasets indicates that the BF is not a result of systematic errors.

Thus when considered by itself, the BF disagrees with $\Lambda$CDM at $\gtwid 98\%$ CL, but there is no disagreement when all 19 moments are considered together. This is a consequence of the fact that we have only one component, the Galactic y-direction of the BF, that is much higher than expected. When considering only three components of the BF, then clearly if one disagrees, the $\chi^2$ will be much higher (probability much lower) than if only one out of nineteen amplitudes is in conflict. This trend is clearly shown in Table~\ref{tab:Pchisq}.

On the other hand, as noted above, for the shear and octupole moments, the measurement uncertainties are greater than for the BF because these moments are essentially derivatives of the (noisy) flow field.  Thus the shear, for example, is not as powerful a cosmological probe as the BF, at least with current datasets. 
One way to quantify this is to compare the WMAP5 CRMS to the measurement noise (see Figs.~\ref{fig:flows-bf-sh}, \ref{fig:flows-octupole} and Tables~\ref{tab:values-19}, \ref{tab:values-9}).  In the case of the BF, this ratio is in the range $1.5-2.5$, whereas for the shear it's $1-1.2$ and for the octupole the values are only $0.5-1$. Thus while we have shown that these higher moments are not much different than expected, due to the noise we do not have much of a handle on their actual values.   Consequently, the discriminatory power of these higher order moments (and their $\chi^2$ values) is less than for the BF, since the contribution to the $\chi^2$ from measurement-noise-dominated moments is always going to be about unity.

Finally, note that the shear tensor tabulated in Tables~\ref{tab:values-19} and \ref{tab:values-9} and shown in Figure~\ref{fig:flows-bf-sh} has 6 components. In practice, however, the trace of the shear (i.e. the average expansion rate) is typically not measurable by peculiar velocity surveys, but is a free parameter that is usually adjusted to be close to zero by requiring that there be no net inflow or outflow of peculiar velocity tracers. Therefore, it is more conservative to consider the traceless components of the shear. If we subtract off 1/3 of the trace from each component on the diagonal ($xx$, $yy$ and $zz$), we see that the measured shears are similar to the WMAP5 expected rms values.

\begin{table*}
\caption {The total observed $P(>\chi^2)$ in percent for \nmom= 3, 9 and 19 and for the BF, shear and octupole moments for each, for $R_I=50$\hmpc, and the WMAP5 central parameters $\Omega_m=0.258$ $\sigma_8=0.796$. } 
\begin{tabular}{lc |||| rrr |||| rrrrr}

& $N_{\rm MOM}=3$ & \multicolumn{3}{c}{$N_{\rm MOM}=9$} && \multicolumn{4}{c}{$N_{\rm MOM}=19$} \\
& BF & Total & BF & shear & & Total & BF & shear & octupole \\

\hline \hline
  COMPOSITE &     1.89 &   6.01 &     1.81 &    41.76 & & 17.00 &     0.50 &    52.60 &    78.33  \\\hline 
 SFI++ &     3.11 & 1.73 &     3.22 &     7.70 & &16.19 &     0.22 &    11.22 &    89.38  \\\hline 
 DEEP &     6.02 &  30.41 &     6.29 &    82.62 & &55.54 &     3.18 &    91.22 &    81.61  \\\hline 
\hline
\label{tab:Pchisq}
\end{tabular}
\end{table*}

\section{Discussion}
\label{sec:discussion}

Our total BF, calculated for the COMPOSITE catalogue at $R_I=50$ \hmpc, with \nmom=19 gives $|v|= 416 \pm 78$ km/s towards Galactic $l = 282^\degree \pm 11^\degree$ and $b =   6^\degree \pm  6^\degree$ which is in disagreement with the expectations of the WMAP5 \citep{WMAP5} cosmology at the 98-99.5\% CL. This result, however, is in excellent agreement with the results found previously (Paper I) estimating only the BF.  The small magnitude of the shear and octupole moments suggests that our previous result was not due to, for example,  aliased small-scale power contaminating the BF measurement. This shows that our orthogonality procedure (Eq.~\ref{eq:mode-fn}) works very well.

This BF value also agrees remarkably well with other peculiar velocity estimates \citep[e.g.][]{Hud94,HofEldZar01,ZarBerdaC01,PikHud05,SarFelWat07,FelWat08,LavTulMoh10}, though these estimates were derived with different catalogues, methodologies and assumptions. 

The results for the shear and octupole are consistent with the hypothesis that the power is not unusual on scales smaller than the very large ones probed by the BF. As mentioned above, another way to make such a comparison is via supercluster infall. For example, \citet{PikHud05} found $(\Omega_{\mathrm m}/0.3)^{0.55} \sigma_{8} = 0.80\pm0.05$, consistent with the mean WMAP5+BAO+SN values $(\Omega_{\mathrm m}/0.3)^{0.55} \sigma_{8} = 0.77\pm 0.035$ \citep{KomDunNol09}.  \citet{AbaErd09}, who studied the velocity correlation function of SFI++, a statistic that mixes a range of scales, also found consistency with WMAP5 cosmological parameters. Thus it does seem to be the BF which is unusual.

Recently, \cite{KasAtrKoc08,KasAtrKoc10} and \citet{AtrKasEbe10} claimed to have detected a dipole in filtered WMAP CMB temperature maps measured at the locations of rich clusters. The magnitude of this cluster-temperature dipole is 2.8 $\pm$ 0.7 $\mu$K. The authors interpret this as being due to the kinetic Sunyaev-Zel'dovich (kSZ) effect, implying a large scale flow with a bulk velocity in the range $\in [600,1000]$ km/s towards $l=283^\degree\pm14^\degree$, $b=12^\degree\pm14^\degree$. Although the scale of their cluster catalogues ($z\sim0.1-0.2$) is of order three or more times the scales that our catalogues probe, this kSZ flow is in excellent directional agreement with the one we detected here and in Paper I. The magnitude is larger, but the authors caution that the magnitude is systematically uncertain.  We note that the BF that we detect does not seem to level off yet and there is no clear sign of convergence to the CMB frame in our catalogues (see Figure~\ref{fig:flows-bf-sh}, top center panel). We by no means claim to verify the \citet{KasAtrKoc08,KasAtrKoc10} results, however, we certainly can not refute it with our data.

The motion we detect is not due to nearby sources, such as the Great Attractor (distance of $\sim40$\hmpc), but rather to sources at greater depths that have yet to be fully identified. The largest known mass concentration, the Shapley supercluster, does not seem to be massive enough to cause a flow of this magnitude \citep{Ray89}. It is likely that the flow arises both from various mass concentrations in the Galactic y-direction as well as under-dense regions in the opposite direction. Currently, there is no peculiar velocity survey in existence that is deep enough to resolve the source(s) of the flow, if indeed it is a potential flow.  

\begin{figure}
\includegraphics[width=\columnwidth]{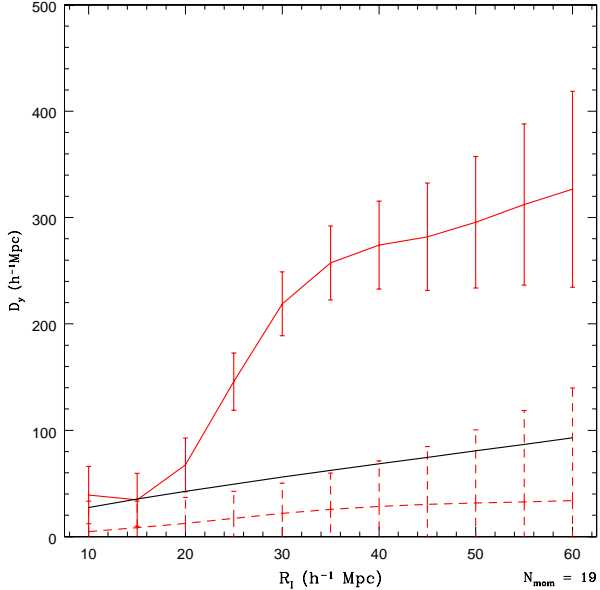}
\caption{\label{fig:attractor}  
The expected distance to the dominant gravitational sources (solid line) ($u_y/u_{yy}$) as a function of $R_I$ for the COMPOSITE survey as well as an expected distance using the WMAP5 central parameters (solid line, no errorbars) discussed in the paper. For comparison we also show $u_x/u_{xx}$  (dashed line) that agrees with the WMAP5 expectations within errors.
}
\end{figure}

A good measure for the distance scale to the sources responsible for most of the BF is the ratio of the BF to shear \citep{LilYahJon86,Kai91}. This characteristic depth $D_{i}\propto u_i/u_{ii}$ with constant of proportionality of order one should be valid even if multiple sources, including underdense regions, as expected for a Gaussian random field, are responsible for the flow.  In Fig.~\ref{fig:attractor} we show the effective distance $D_{y}\propto u_y/u_{yy}$ to an attractor for the COMPOSITE survey as a function of $R_I$. Nearby, where $R_{I} \ltwid 30$ \hmpc,  the moments are dominated by the nearby velocity field and show the effect of the Great Attractor at $\sim 35$ \hmpc. Then there is a sudden rise around 30\hmpc\ that suggests that on scales beyond the Great Attractor, the flow is driven by even larger and more distant mass concentration(s) in the same general direction. The uncertainties are large, however, at $R_I=50$\hmpc\  we find a characteristic distance of $296\pm 62$ \hmpc, much larger than the expected WMAP5 value of 82 \hmpc. For comparison we show the estimated $u_x/u_{xx}$ which is consistent with the WMAP5 expected results. 

Note, however, as discussed above, it is sensible to subtract 1/3 of the trace of the shear tensor from each diagonal component. In this case, the $yy$ component of the shear tensor is consistent with zero, implying that the attractor is at infinity.  Thus we cannot exclude the coherent flow on much larger scales claimed by \citet{KasAtrKoc08, KasAtrKoc10}. 

\section{Conclusions}
\label{sec:conclusions}

We have estimated the MV BF, shear and octupole moments and showed that they are minimally sensitive to  aliasing from small-scale power. We have applied the MV formalism to a number of compilations of recent peculiar velocity surveys. The MV window functions are similar to the ideal window functions and our octupole moments orthogonalization procedure works well in a sense that including them does not affect the BF much.

The various peculiar velocity surveys are consistent with each other and all show a large BF which does not follow expectations from the WMAP5-normalized $\Lambda$CDM model. Specifically, we have shown that the BF within a Gaussian window of radius $R_I=50$\hmpc\ has a magnitude of  $416 \pm 78$ km/s towards Galactic $l = 282^\degree \pm 11^\degree$ and $b =   6^\degree \pm  6^\degree$  in disagreement with WMAP5 \citep{WMAP5} at the 99.5\% CL.  This flow is consistent with being relatively ``cold'' (both shear and octupole moments are in agreement with expectations) and is not due to nearby sources. If we include all nineteen octupole (nine shear) moments, we disagree with the WMAP5 expectations at only the 83\% (94\%) CL, respectively. We also found that the sources responsible for the BF are at an effective distance of $> 300$ \hmpc, too far to identify in existing all-sky redshifts surveys. 

If the flow is a potential flow, there are various possible explanations as to its source. While it is always possible that the result is due to systematic errors in the data, we tend to discount this possibility since the BF is seen in many independent surveys with various distance indicators, methodologies and geometries. It may be that we live in a $\Lambda$CDM Universe but that we happen to live in one of the very rare volumes that exhibits this flow ($\ltwid1$\%), or we may live in a Universe with more large-scale power than WMAP5-normalized $\Lambda$CDM.  In either of these last two cases, we may be able to find the source or sources of the flow in future redshift surveys. It is of course likely that the sources are a combination of over- and under-dense regions and not just a single large mass concentration. 

\vskip 0.1cm
\noindent{\bf Acknowlegements:} 
We thank our referee, Michael Strauss, for useful comments.
HAF has been supported in part by an NSF grant AST-0807326, by a grant from the Research Corporation, by the National Science Foundation through TeraGrid resources provided by the NCSA, by the University of Kansas General Research Fund (KUGRF) and acknowledges the hospitality of University College, London and Imperial College in the UK and the Institut d'Astrophysique de Paris, France.  MJH has been supported by NSERC and acknowledges the hospitality of the Institut d'Astrophysique de Paris, and the financial support of the IAP/UPMC visiting programme and the French ANR (OTARIE).

\bibliographystyle{mn2e}
\bibliography{ms}

\end{document}